\documentstyle[floats,psfig,prc,aps,manuscript]{revtex}

\begin{document}

\title{ 
Squeeze-out of nuclear matter in peripheral heavy-ion collisions
and momentum-dependent effective interactions.
\footnote{Supported by BMBF and GSI Darmstadt} 
      } 

\author{ A. B. Larionov \footnote{On leave from RRC 
         "I.V. Kurchatov Institute", 123182 Moscow, Russia}, W. Cassing,
         C. Greiner, U. Mosel }
         
\address{ Institut f\"ur Theoretische Physik, Universit\"at Giessen,
          D-35392 Giessen, Germany }

\maketitle

\begin{abstract}
We perform a systematic study of in-plane and out-of-plane proton and neutron
flow from nucleus-nucleus collisions within the BUU transport approach 
employing different parameter sets for the mean-field potentials 
that can be characterized by the nuclear incompressibility and the
stiffness of the momentum-dependent potential.  
We find that a simultaneous description of the experimental data from the
BEVALAC and the SIS on both the nucleon squeeze-out $v_2$ and
the in-plane flow $F$ at beam energies $E_{lab} = 0.15 \div 2$ AGeV 
requires a mean field with strong momentum dependence,
i.e. an effective Landau mass $m^* \simeq 0.68 m_0$
at normal nuclear matter density $\rho_0=0.17$ fm$^{-3}$, 
where $m_0=0.938$ GeV is the bare nucleon mass. Some experimental data on 
the squeeze-out require an even stiffer momentum dependence 
($m^* \approx 0.62 \ m_0$). All systems investigated are found to be
compatible with $m^*/m_0 = 0.65 \pm 0.03$.
\end{abstract}

\vspace{0.5cm}

\hspace{-\parindent}

\hspace{-\parindent}
Effective mass 

\newpage

\section{ Introduction }

Since many years the study of the various collective flows of nuclear 
matter is one of the main subjects in heavy-ion physics 
\cite{Scheid68,SG86,BD88,Gut89,Aich91,Reis97,Sahu98,Herrm99,Hombach99}. 
The interest in collective nuclear motion under extreme conditions, 
like high density and/or high temperature, originates from the fundamental 
problem to extract the equation of state (EOS) of nuclear matter. 
It was shown in Refs. \cite{GBD87,Aich87,Gale90,Zhang94} that the in-plane 
directed flow \cite{DO85} is sensitive to both the incompressibility $K$ of 
the EOS and to the momentum-dependent effective interaction (MDI). 
Moreover, the in-medium nucleon-nucleon (NN) cross section (that is related 
to the imaginary part of the in-medium $G$-matrix) also influences the 
transverse directed flow \cite{balance} since nucleons acquire transverse 
momentum also by elastic and inelastic collisions. It is necessary, therefore,
to have at least two additional observables, besides the in-plane flow, in 
order to determine the three most commonly used components of the effective 
NN interaction adopted in the various analyses: the incompressibility $K$, 
the stiffness of the MDI and, possibly, the in-medium NN cross section.
  
The rapidity distribution of nucleons, which is only weakly dependent on 
the mean-field parameters, gives reasonable constraints on the in-medium 
NN cross section or on the nuclear stopping power \cite{Victor,Homb99}. 
However, it was not so clear until recently, how to possibly extract 
information on the MDI from heavy-ion collisions. Danielewicz \cite{Dan99}
has shown within the Boltzmann-Uehling-Uhlenbeck (BUU) transport model, 
that the elliptic flow (or squeeze-out ratio) of protons at midrapidity 
is very sensitive to the MDI part of the nucleon mean field. 
The reaction Bi+Bi at $E_{lab}$=400, 700, 1000 AMeV (measured by the KAOS 
Collaboration \cite{Brill96}) has been analysed in \cite{Dan99} leading to 
the conclusion that the MDI part of the NN interaction should produce an 
effective Landau mass $m^* \simeq 0.7 m_0$, since momentum independent forces 
($m^*=m_0$) strongly underpredict the squeeze-out ratio at high transverse 
momenta \cite{Dan99}.

The aim of our present work is: (i) to study the mechanism of the 
squeeze-out enhancement by the MDI and (ii) to analyse several independent 
data sets on the {\it directed} and {\it elliptic} flow of nucleons to get 
experimental constraints on the stiffness of the MDI. We concentrate
on the bombarding energy regime up to 2 AGeV since here multiple meson
production or "string" degrees of freedom play a minor role. As shown 
by Sahu et al. \cite{Sahu2000} the collective flow is strongly influenced
by the effective degrees of freedom taken into account in the transport
approach at high energies. For a detailed investigation of this problem
-- especially at AGS energies -- we refer the reader to Ref. \cite{Sahu2000}.
 
The paper is structured as follows: In Sect. II a brief description
of the applied BUU model is given. Sect. III contains our interpretation
of the squeeze-out mechanism which is illustrated in Sect. IV by 
a detailed study of the time evolution of colliding nuclear systems within 
BUU calculations. 
A comparison to the experimental data on the proton and neutron elliptic 
and directed flows is given in Sect. V, while Sect. VI concludes our work 
with a summary of the results. 

\section{ The BUU model }

In our calculations we have applied the recent version of the BUU model 
described in more detail in Ref. \cite{EBM99}. The equation of motion for the 
nucleon phase-space distribution function $f({\bf r},{\bf p},t)$ includes a 
propagation in the momentum-dependent scalar mean field $s({\bf r},{\bf p},f)$ 
via the single-particle energy
\begin{equation}
         H_{mf}({\bf r},{\bf p},t) = 
         \sqrt{ (m_0+s({\bf r},{\bf p},f))^2 + p^2 }~.      \label{Hmf}
\end{equation}
Both elastic and inelastic scattering processes are additionally described 
by the r.h.s. of the transport equation 
\begin{eqnarray}
\lefteqn{  \left( {\partial \over \partial t} + 
            {\partial H_{mf} \over \partial {\bf p}_1}
            {\partial \over \partial {\bf r}} -
            {\partial( H_{mf} + U_{Coul} ) \over \partial {\bf r}}
            {\partial \over \partial {\bf p}_1}
           \right) f({\bf r},{\bf p}_1,t) }  & &              \nonumber \\
 & = & g \int { d{\bf p}_2 \over (2\pi\hbar)^3 } \int d\Omega\, v_{12} 
    { d \sigma_{NN}(p_{12}) \over d \Omega }
    ( f_3f_4(1-f_1)(1-f_2) - f_1f_2(1-f_3)(1-f_4) )           \nonumber \\
 & + & {\rm coupling~terms}~,                                 \label{BUUeq} 
\end{eqnarray}
where $U_{Coul}$ is the Coulomb potential acting on protons;
$g=4$ is the spin-isospin degeneracy of a nucleon; 
$v_{12}=|{\bf v}_1-{\bf v}_2|$ and  $p_{12}=|{\bf p}_1-{\bf p}_2|$
are the relative velocity and relative momentum of colliding nucleons,
respectively; 
$d \sigma_{NN}(p_{12}) / d \Omega$ is the energy dependent
differential NN scattering cross section as parametrized by Cugnon 
\cite{Cugnon96};
$f_i := f({\bf r},{\bf p}_i,t)$ $(i=1,2,3,4)$, while ${\bf p}_3$ 
and ${\bf p}_4$ are the final momenta of the scattered nucleons.
The coupling terms in (\ref{BUUeq}) account for the various 
inelastic channels:
$N N \Leftrightarrow \pi N N$, $N N \Leftrightarrow N R$ 
-- S-wave pion and resonance production/absorption, 
$N N \rightarrow  \Lambda K N$, $N N \rightarrow  \Sigma^0 K N$ 
-- strangeness production.

The scalar potential $s({\bf r},{\bf p},f)$ is computed as follows
(see \cite{Teis97}):\\
1) In the local rest frame (l.r.f.) of the nuclear matter element, where 
the space components of the baryonic 4-current $j^\mu=(j^0,{\bf j})$
vanish, i.e. ${\bf j} = 0$, the single-particle energy is calculated as
\begin{equation}
\epsilon_{l.r.f.}({\bf r},{\bf p}) 
= \sqrt{ m_0^2 + p^2 } + U(\rho,{\bf p})~.
                                                       \label{epslrf}
\end{equation}
The explicit form of the momentum-dependent potential $U$ 
in (\ref{epslrf}) is taken from Welke et al. \cite{Welke88} as
\begin{equation}
U(\rho,{\bf p}) = A { \rho \over \rho_0 } + 
B\left({ \rho \over \rho_0 }\right)^\tau +
{ 2C \over \rho_0 } \int { gd{\bf p}^\prime \over (2\pi\hbar)^3 }
{ f({\bf r},{\bf p}^\prime) \over 
  1 + ({\bf p}-{\bf p}^\prime)^2/\Lambda^2 }~.
                                                       \label{U}
\end{equation}
A symmetry energy term is not taken into account in the potential (\ref{U}).\\
2) The mean field $s$ is calculated from the relation in the l.r.f.
\begin{equation}
s = \sqrt{ \epsilon_{l.r.f.}^2 - p^2 } - m_0~.           \label{epslrf1}
\end{equation}
3) The scalar potential $s$ is then used in Eq. (\ref{Hmf}) to determine
the single-particle energy in the computational frame.
Steps 1)-3) are repeated until selfconsistency is reached, i.e.
$s$ is not changed within the accuracy of numerics on the level of
$10^{-4}$.

The interaction range $\Lambda$ of the potential (\ref{U}) is chosen
as in Ref. \cite{Welke88}: $\Lambda = 1.5 p_F^{(0)}$, where
$p_F^{(0)} \equiv p_F(\rho_0)$,
$p_F(\rho) = ({3 \over 2} \pi^2 \rho)^{1/3} \hbar$.
The residual four free parameters $A$, $B$, $\tau$ and $C$ of the potential
$U$ are determined from the conditions:\\
(i) The effective mass at the Fermi surface 
\begin{equation}
(m^*)^{-1} = m_0^{-1} + (p_F^{(0)})^{-1} 
{ \partial U \over \partial p }_{|p=p_F^{(0)}}              \label{meff}
\end{equation}
must be in the range $0.6m_0 \leq m^* \leq m_0$ 
(c.f. \cite{Prakash97,Chab97}). This condition puts limits on the stiffness
$C$ of the MDI.\\ 
(ii) The energy per nucleon in nuclear matter has a minimum at 
$\rho=\rho_0$ and\\ 
(iii) a value of $-16$ MeV at the minimum:
\begin{equation}
{ \partial \epsilon/\rho \over \partial \rho }_{|\rho=\rho_0} = 0~,~~~~~
{ \epsilon \over \rho }_{|\rho=\rho_0} = -16~\mbox{MeV}~,     \label{satur}
\end{equation}
where 
\begin{eqnarray}
\epsilon(\rho) &=& { 3 \over 5 } \epsilon_F \rho + 
{ 1 \over 2 } A { \rho^2 \over \rho_0 } + 
{1 \over \tau + 1 } B { \rho^{\tau+1} \over \rho_0^\tau } +  \nonumber \\
& & { C \over \rho_0 } \int { gd{\bf p}_1 \over (2\pi\hbar)^3 }
{ gd{\bf p}_2 \over (2\pi\hbar)^3 }
{ f({\bf p}_1) f({\bf p}_2) \over 
  1 + ({\bf p}_1-{\bf p}_2)^2/\Lambda^2 }                    \label{edens}
\end{eqnarray}
with $\epsilon_F = p_F^2/2m_0$ and $f({\bf p})=\Theta(p_F-p)$. Explicit forms
of the momentum integrals in (\ref{meff}),(\ref{edens}) are given in
Ref. \cite{Welke88}.\\ 
(iv) The nuclear matter incompressibility  
\begin{equation}
K = 9 \rho_0^2 
{ \partial^2 \epsilon/\rho \over \partial \rho^2 }_{|\rho=\rho_0}
                                                             \label{compr}
\end{equation}
is fitted in the range $K=200 \div 380$ MeV.

We have applied in our calculations the following mean fields denoted
by H (hard, without momentum dependence), HM and SM (hard and soft, with 
a "standard" momentum dependence) as well as the new mean-field 
parametrization which is dubbed HM1 (see below).
The parameters of all interactions are presented in Table 1.
The stiffness $C$ of the MDI in the interactions HM and SM is 
close to that of Ref. \cite{Welke88}. The set of parameters HM1 
includes an MDI stiffness that is about 30\% larger than in HM and SM, 
which leads to a smaller effective mass at the Fermi momentum 
for HM1. In Fig. 1 we show the 
resulting energy per nucleon as a function of density and the potential 
$U(\rho,p)$ versus momentum for several densities. 
The interactions H, HM and HM1 give practically the same EOS
($E/A(\rho)$-dependence), but they differ in the MDI part, i.e. in their 
momentum dependence. This gives the possibility to extract an independent 
information on the MDI stiffness $C$. On the other hand, the interactions 
HM and SM produce different EOS, but have the same MDI, which is
relevant for a separate study on the sensitivity to the incompressibility 
$K$. 

\section{ Squeeze-out mechanism }

A squeeze-out of nucleons is observed experimentally in the beam energy 
range below 1$\div$2 AGeV where mean-field effects still play an essential 
role. It can be interpreted, however, to a major extent in terms 
of a shadowing phenomenon \cite{Aich91}, as in the participant-spectator 
picture of a heavy-ion collision the fireball particles -- 
emitted in the reaction plane -- are rescattered on the spectators. 
This results intuitively in a squeeze-out ratio 
$R_N := (N(90^o)+N(270^o))/(N(0^o)+N(180^o))$ at midrapidity being larger
than 1. Here $N(\phi)$ is the azimuthal distribution of nucleons with 
respect to the reaction plane in a given rapidity interval. 
A specific particle emitted from the center-of-mass of a 
system and moving in transverse direction with velocity $v_t$ will be 
shadowed by the spectator piece, if it reaches the spectator during the 
passage time $\Delta t = 2R/(v_p^{c.m.}\gamma)$ of the colliding nuclei. 
Here $R$ is the radius of the nuclei (assummed to be equal for projectile
and target), $v_p^{c.m.}$ 
is the projectile velocity in the center-of-mass system, and  
$\gamma=(\sqrt{1-(v_p^{c.m.})^2})^{-1}$. A straightforward geometrical
estimate then leads to the condition $v_t\Delta t/2 > R - b/2$, or
\begin{equation}
v_t > { R - b/2 \over R } \gamma v_p^{c.m.}~,           \label{shadcond}
\end{equation}
where $b$ is the impact parameter.
One can see from (\ref{shadcond}) that at larger impact parameters
$b$ the lower limit of $v_t$ becomes smaller, i.e. more particles will
be shadowed and the squeeze-out ratio $R_N$ will increase
(c.f. later Fig. 12).
This is simply related to a larger size ($\sim b$) of the spectator pieces in
peripheral collisions. It is also evident, that for faster moving particles
it is easier to fulfill the condition (\ref{shadcond}) and, therefore,
the squeeze-out ratio should grow with the particle transverse velocity
$v_t$ (c.f. later Figs. 6,13).

The squeeze-out phenomenon can also be characterized by a negative elliptic 
flow $v_2$
(c.f. \cite{App98}) which appears in the Fourier expansion
\begin{equation}
N(\phi) \propto 1 + 2v_1\cos(\phi) + 2v_2\cos(2\phi) + ...~.   \label{Fourier}
\end{equation}
Neglecting thus higher order terms in (\ref{Fourier}), the following 
direct relation is obtained:
\begin{equation}
R_N \simeq { 1 - 2v_2 \over 1 + 2v_2 }~.                    \label{RNv2}
\end{equation}
In (\ref{Fourier}) the elliptic flow $v_2$ is given as
\begin{equation}
v_2 = \left\langle { p_x^2 - p_y^2 \over p_x^2 + p_y^2 } \right\rangle ~. 
                                                               \label{v2}
\end{equation}

At higher beam energy ($\sim 10$ AGeV), faster moving and Lorentz-contracted
spectators do not shadow  particles any more from the expanding fireball
\cite{Sahu2000}. This results in $R_N \leq 1$ (or $ v_2 \geq 0 $) as observed
in experiment \cite{E895}.

The simple geometrical picture discussed above will be illustrated by the
BUU phase-space evolution in the next section. Since the squeeze-out 
is related to the participant-spectator reaction mechanism 
(c.f. \cite{Gosset77}), we expect a lower beam-energy limit of
$\sim 0.1$ AGeV for $R_N > 1$. At even lower beam energies the mean-field
effects lead to the formation of rotating nuclear systems \cite{Tsang96}
that emit particles predominantly in the reaction plane due to centrifugal
forces, thus producing $R_N < 1$ (or $v_2 > 0$, c.f. later Fig. 7).  

We have to note, however, that the shadowing scenario can not completely
explain the squeeze-out phenomenon. For instance, a small but statistically
significant increase of $-v_2$ at the beam energy $\sim 0.4$ AGeV
(see later Fig. 7) with increasing incompressibility $K$ can not be
understood in the shadowing picture. This is, probably, a manifestation
of a focussing of high momentum particles by the repulsive mean 
field as proposed in Ref. \cite{Dan99}.

\section{ BUU study of the squeeze-out time evolution }

We study the system Au+Au at the beam energy $E_{lab}=0.4$ AGeV
and impact parameter $b=6$ fm. Fig. 2 shows the time evolution of
the central baryon density and of the NN collision rate for the HM 
and SM mean fields. In agreement with Fig. 1 (upper left panel), a slightly
higher density and, as a consequence, a higher collision rate are reached 
in the case of the SM parametrization. However, the various EOS shown
in Fig. 1 practically do not differ in the density interval  
$0< \rho \leq 0.27$ fm$^{-3}$ probed in the collision. We do not expect,
therefore, to get clear constraints on the incompressibility $K$ from
our study of {\it peripheral} collisions and concentrate mainly on 
the MDI-part of the effective interactions.  

As pointed out in Ref. \cite{Dan99} the $p_t$-dependence of the $v_2$ 
coefficient is very sensitive to the momentum dependence of the nuclear 
mean field. In order to understand this fact, we have performed calculations 
with the parametrizations H and HM of the mean field.
Fig. 3 shows the average transverse velocity (a) and transverse momentum
(b) of the midrapidity neutrons moving in- and out- of the reaction plane 
versus time. We observe that in the case of the HM parameter set 
$\langle v_t \rangle$ reaches a maximum at $t \simeq 17$ fm/c when also
the central density and the collision rate are close to the maximum (Fig. 2).
At this time the system shows the most compact configuration in space, 
i.e. the line connecting the centers of target and projectile spectators,
is perpendicular to the beam (z-) direction (Fig. 4). However, in case of the
H mean field the transverse velocity has no maximum (the peaks
of $\langle v_t \rangle$ and $\langle p_t \rangle$ at $ t \simeq 7$ fm/c
are due to fluctuations, because initially there are no nucleons at 
midrapidity). 
Moreover, at $t < 40$ fm/c in the case of the HM mean field the average
transverse velocity is larger than for the mean field H, while 
the transverse momentum is practically independent on the mean field, even 
showing a slightly opposite tendency: at $t < 40$ fm/c the mean field H 
produces a somewhat larger $\langle p_t \rangle$. This can be qualitatively 
understood from the Hamiltonian equations:
\begin{eqnarray}
{\bf \dot r} & = & { \partial H_{mf} \over \partial {\bf p} }
= { {\bf p} \over H_{mf} } 
+ { m_0 + s \over H_{mf} } { \partial s \over \partial {\bf p} }~, 
                                                            \label{vel}\\
{\bf \dot p} & = & -{ \partial H_{mf} \over \partial {\bf r} }
= - { m_0 + s \over H_{mf} } { \partial s \over \partial {\bf r} }~.
                                                            \label{acs}
\end{eqnarray}
where $H_{mf}$ is the single-particle energy (\ref{Hmf}). For the 
momentum-independent mean field H, (\ref{vel}) reduces to a simple 
proportionality: ${\bf v} = {\bf p}/H_{mf}$, where $H_{mf} \sim 1$ GeV. 
However, for the HM interaction the strong momentum-dependent mean field 
$s({\bf r},{\bf p},f)$ is created by the density build-up, and the second 
term in the r.h.s. of (\ref{vel}) becomes large (and positive), which 
results in a larger transverse velocity for HM.
On the other hand, the force acting on a particle depends mostly on the EOS 
produced by a given mean field or on the pressure gradient. Thus, the r.h.s.
of (\ref{acs}) depends relatively weakly on the momentum dependence of a 
mean field. 

Later on $\langle v_t \rangle$ and $\langle p_t \rangle$ decrease reaching 
asymptotic values at $t \simeq 50$ fm/c. This decrease is due to less 
energetic products of NN scattering, which populate the midrapidity region 
at later times, while first-chance collision products have larger kinetic 
energies. 
Faster moving particles (in the case of the HM calculation with respect 
to H) are shadowed more effectively, which causes a larger splitting between 
in- and out-of-plane transverse velocities for the HM calculation 
at large times.

In Fig. 5 we show the final azimuthal angle dependence of the midrapidity 
neutron transverse velocity (a) and the neutron azimuthal distribution (b).
The azimuthal angle modulation of the transverse velocity is clearly
visible, which was first observed experimentally \cite{Rai99}.  
The transverse velocity azimuthal angle dependence and the azimuthal 
distribution of particles have the same shape. 
The depletion of the particle yield at $\phi = 0^o$ and $180^o$ is
caused by a shadowing of fast moving particles by the spectators;
the enhanced yield at $\phi = \pm 90^o$ correlates with the
nonshadowed emission of fast particles.

Fig. 6 shows the coefficient $v_2$ vs. transverse velocity (left panels)
and vs. transverse momentum (right panels) at the two time steps
$t = 30$ fm/c and $t = 50$ fm/c. Even at 30 fm/c the particles still feel
the influence of the nuclear field and, therefore, the simple relation
$v_t = p_t/H_{mf}$ can be applied at this moment only in the case of
the momentum-independent mean field H. One can see, that at 30 fm/c 
the $v_2(p_t)$-dependence is different for different mean fields, while 
the $v_2(v_t)$-dependence is very similar for the H and HM calculations.
This is in agreement with our expectation, that particles moving with
the same transverse velocity are shadowed in the same way regardless of
their transverse momenta. At later times, this "transient universality"
of $v_2(v_t)$ is destroyed since the influence of the mean field will
become negligible here and $v_t = p_t/\sqrt{m_0^2+p^2}$ for both 
parameter sets.

\section{ Comparison with experiment }

In our comparison to data on squeeze-out observables we now 
compare in parallel also to data on the in-plane flow $F$, which
is defined as the derivative of the transverse in-plane momentum
component $\langle p_x \rangle$ with respect to the normalized c.m. 
rapidity $y^{(0)}=(y/y_{proj})_{c.m.}$ at midrapidity \cite{Doss86}:
\begin{equation}
      F = { d\langle p_x \rangle \over dy^{(0)} }_{|y^{(0)}=0}~.     
                                                  \label{flowdef}
\end{equation}  
A simultaneous description of both observables, i.e. $v_2$ and $F$, 
will give a rather strong confidence that the dynamics of the participant 
zone and the interaction of the participating nucleons with the spectators 
are described consistently in our calculation and potentially leading to 
stringent constraints on the EOS and the interactions employed (see also 
Refs. \cite{Zheng99,Soff} where the energy regime below 100 AMeV has been 
studied in detail).

Figs. 7-9 show the excitation functions of $v_2$  and
$F$ for protons in case of Au+Au collisions in the impact 
parameter range $b=5\div7$ fm corresponding to an estimate given in 
Refs. \cite{E895,Part95}. The data are taken from Refs. 
\cite{Herrm99,Part95,Liu98,Liu00,Mag00}.
The time evolution has been followed until $t_{max}$=40 fm/c at
$E_{lab}$=2 AGeV, while at smaller beam energies $t_{max}$ gradually
increases reaching 250 fm/c at $E_{lab}$=25 AMeV.

The $v_2$ coefficient (Fig. 7) was calculated for {\it free} protons 
in the c.m. rapidity range $|y| < 0.1$.
Free protons were selected by the requirement, that the distance
to the closest particle within a given parallel ensemble is more
than some critical distance $d_c \simeq 3$ fm in order to separate
from protons bound in fragments. 
As discussed in Sect. III, the elliptic flow $v_2$ 
vs. $E_{lab}$ shows a nonmonotonic behaviour: it decreases from positive 
to negative values reaching a minimum at $E_{lab} \simeq 0.4$ AGeV, 
and then it starts to increase moderately again with beam energy. 
In this work, however, we only concentrate on the beam energy domain 
below 2 AGeV and, therefore, the transition from negative to positive 
$v_2$ at a beam energy $\sim 4$ AGeV \cite{E895,Sahu2000} is not discussed 
here. The lower transition energy, i.e. the beam energy at which $v_2$ 
changes sign, is $E_{TRA} \simeq 100$ AMeV, in agreement with the FOPI-data
\cite{Crochet97}. The in-plane flow (Figs. 8,9) increases monotonically in 
the beam energy region considered. The balance energy, i.e. the beam 
energy at which the flow $F$ changes sign, is $E_{BAL}\simeq 50$ AMeV
in our calculation. 
This value is in between the MSU-data \cite{Mag00} ($E_{BAL}\simeq 
42$ AMeV) and the FOPI-data \cite{Crochet97} ($E_{BAL}\simeq 60$ AMeV). 

Now we discuss the effect of the mean field on $v_2$ and $F$. 
The calculation with a momentum-independent mean field H is in a 
satisfactory agreement with proton in-plane flow data below 1 AGeV (Fig. 8),
but it fails to reproduce the strong squeeze-out at $E_{lab} \simeq 0.4$ 
AGeV (Fig. 7). The HM parametrization increases the in-plane flow and 
squeeze-out (see also Fig. 5) and, thus, the in-plane proton flow $F$ 
is now overpredicted. The lower incompressibility $K$ in the SM calculation 
results in a reduced squeeze-out and a reduced flow with respect to the 
HM calculation. Generally, the in-plane proton flow and $v_2$ data 
are best reproduced with the SM mean field.

In Fig. 9 we compare to FOPI, EOS and recent MSU-$4\pi$ data on the in-plane
flow excitation function. The selected data sets are now for a mixture of 
protons and complex fragments and, therefore, they exhibit a stronger
flow signal (c.f. Fig. 8, where the data are for protons only). A separate
study of the fragment flow is out of scope of this work. However, in order
to understand qualitatively the possible effects from cluster formation,
we present in Fig. 9 the results of the BUU-SM calculation both for
{\it all} (solid line) and {\it free} (dashed line) protons. We find a 
steeper increase of the in-plane flow for {\it all} protons with the beam 
energy from 0.1 to 1 AGeV. The calculation for {\it all} protons is 
in a good agreement with data at $E_{lab}=0.4\div1$ AGeV, but it
underpredicts the flow at lower beam energies, where the {\it free}
proton flow $F$ increases earlier with $E_{lab}$ and is closer to the data.
This can be explained by the cluster formation scenario from initially 
{\it free} emitted protons by the momentum (rather than coordinate) space 
coalescence.
The balance energy given by the SM calculation is $E_{bal} \simeq 
50$ (60) AMeV for {\it free} ({\it all}) protons, which is somewhat
larger than the value $E_{bal} \simeq 42$ AMeV reported in \cite{Mag00}.
But the shape of the in-plane flow excitation function at 
$E_{lab}=25\div60$ AMeV is in a remarcable agreement with the 
MSU-$4\pi$ data.
We would like to stress, that the negative in-plane flow in the Au+Au
system below $E_{BAL}$ is obtained in our calculations taking into account 
initial repulsive Rutherford-trajectories (c.f. Ref. \cite{Soff}, where an
opposite result has been obtained). 

In Fig. 10 we present the midrapidity neutron azimuthal distributions 
with respect to the in-plane flow axis for the reactions Nb+Nb and La+La 
at 400 AMeV in comparison to the data from Ref. \cite{Htun99} (see Fig. 11 
of \cite{Htun99}). In this comparison the BUU events were treated as the 
experimental data \cite{Htun99}, i.e. the azimuthal angle of the reaction 
plane was defined by the ${\bf Q}$ vector:
\begin{equation}
{\bf Q} = \sum_\nu w^\nu { {\bf v}_t^\nu \over 
                           |{\bf v}_t^\nu| }~,            \label{Qvec}
\end{equation}
where the sum is taken over all charged fragments (i.e. all protons
in BUU); ${\bf v}_t^\nu$ is the transverse velocity of the particle
$\nu$; $w^\nu=1$ for $y^{(0)} \geq 0.2$, $w^\nu=0$ for 
$-0.2 \leq y^{(0)} \leq 0.2$, $w^\nu=-1$ for $y^{(0)} < -0.2$.
Furthermore, we have determined the flow angle $\Theta_F$ by diagonalizing
the flow tensor $F_{ij}$ in the c.m.s. (see \cite{Gy82}):
\begin{equation}
F_{ij} = \sum_\nu { 1 \over 2m_0 } { v_i^\nu v_j^\nu \over
                                   |{\bf v}^\nu|^2 }~.     \label{flowtens}
\end{equation}
Finally, a rotation of the c.m. coordinate system was performed, first,
around the beam z-axis to align the x-axis with the vector ${\bf Q}$ and,
second, around the y-axis by the flow angle $\Theta_F$ 
(see Refs. \cite{Gutbrod89,Gutbrod90}).
As shown in Refs. \cite{Gutbrod89,Gutbrod90}, the squeeze-out is more 
pronounced in the rotated coordinate system $(x^\prime,y^\prime,z^\prime)$. 
The azimuthal distributions in Fig. 10 are presented for neutrons with 
longitudinal momenta $-0.1 \leq (p_z^\prime/p_{proj}^\prime)_{c.m.} \leq 0.1$.
We see from Fig. 10, that the data can be reasonably well reproduced adopting
the HM1 mean field (see Table 1), suggesting that the MDI stiffness $C$ in 
the parameter set HM is not high enough. Our conclusion is in line with the 
results of Ref. \cite{Htun99}, where it was reported that BUU calculations 
with the mean field of \cite{Welke88} ($K=215$ MeV, i.e. the SM 
parametrization) underpredict the squeeze-out. 

Fig. 11 shows the in-plane transverse momentum $\langle p_x \rangle$ of 
neutrons versus the normalized rapidity $y^{(0)}$ for the reactions Nb+Nb
and La+La at 400 AMeV. We observe that an increased MDI stiffness 
improves the agreement with the neutron in-plane flow data, too.

Finally, we have studied the neutron squeeze-out in Au+Au collisions measured 
by the FOPI-LAND Collaboration \cite{Lambr94}. Fig. 12 shows the impact 
parameter dependence of the squeeze-out ratio $R_N$ for Au+Au collisions
at 400 AMeV, which reasonably describes the data for the HM calculation
(dotted histogram). The HM1 parameter set (solid histogram) underpredicts 
somewhat the squeeze-out ratio in central collisions and overpredicts 
it in semiperipheral ($b=7\div8$ fm) collisions. The SM parameter set 
(dash-dotted histogram) and momentum-independent field H (dashed histogram) 
underpredict the data at large impact parameters.  
The calculated $R_N(b)$-dependence has a characteristic 
bell-shape with a maximum at $b \simeq 7$ fm. This agrees with the results 
of Ref. \cite{Crochet97}, where the squeeze-out was studied for charged 
particles. The drop of $R_N$ at large impact parameters can be explained by 
a smaller compression of the participant zone and, therefore, smaller
transverse velocities of particles emitted in peripheral collisions. 

In Fig. 13 the $p_t$-dependence of the ratio $R_N$ is shown for 
collisions of Au+Au at 400, 600 and 800 AMeV. In agreement with the previous 
calculations of Ref. \cite{Dan99} we observe that a lower effective mass, 
i.e. a steeper momentum dependence, increases the squeeze-out ratio
$R_N$ at large $p_t$ (c.f. dotted and solid lines in Fig. 13 d-f).
On the other hand, reducing the NN cross section decreases the ratio 
$R_N$ at large $p_t$ (c.f. solid and long-dashed lines in Fig. 13 e). 
The parameter set H produces a too flat $R_N(p_t)$-dependence (c.f. 
short-dashed line in Fig. 13 a). The SM calculation gives the best overall 
agreement with the data (dash-dotted line in Fig. 13 a-c). We see that our 
BUU calculations reproduce the experimentally observed trend of $R_N$ to 
decrease (at fixed $p_t$) with collision energy, which was interpreted in 
Ref. \cite{Lambr94} as a scaling behaviour of the squeeze-out.

\section{ Summary and conclusions}

The squeeze-out of nuclear matter in heavy-ion collisions
at beam energies $E_{lab}=0.05\div2$ AGeV has been studied 
within the BUU approach of Ref. \cite{EBM99}. As demonstrated,
being essentially caused by the shadowing of the expanding fireball 
by cold spectators, the squeeze-out is quite sensitive to the specific 
velocity of the particles. A mean field with a larger MDI stiffness gives 
a smaller effective mass, and thus a harder velocity spectrum of nucleons.
In a heavy-ion collision thus faster moving nucleons are shadowed more 
effectively and the squeeze-out ratio $R_N$ is larger for mean fields with 
stronger momentum dependence. 

The systematic comparison with the experimental data on the excitation 
function of the elliptic flow $v_2$ \cite{Herrm99} favours mean fields
with $m^* \simeq 0.68 m_0$, i.e. HM or SM sets. The influence of the 
incompressibility $K$ on the squeeze-out turns out to be quite weak.
The proton in-plane flow data of the EOS Collaboration at 
$E_{lab}=0.25\div1.15$ AGeV and the fragment flow data of the
MSU-$4\pi$ Collaboration at $E_{lab}=25\div60$ AMeV are consistent
with the SM mean field.
We have found, that a reproduction of the BERKELEY data \cite{Htun99} on the 
neutron azimuthal distributions and the in-plane flow in Nb+Nb and La+La 
collisions at 400 AMeV requires an enhanced MDI stiffness 
for $p < 1$ GeV/c (HM1 parameter set: $K=379$ MeV and $m^* = 0.62 m_0$).
However, the FOPI-LAND data \cite{Lambr94} on the neutron squeeze-out in 
Au+Au collisions at 400$\div$800 AMeV favour the standard MDI stiffness 
(HM or SM parameter sets). 

To conclude, a simultaneous description of the nucleon squeeze-out $v_2$ and
the in-plane flow $F$ at the beam energies $E_{lab} = 0.2 \div 2$ AGeV 
requires a mean field with a strong momentum dependence. This corresponds
to an effective Landau mass at the Fermi surface $m^*/m_0 = 0.65 \pm 0.03$
at normal nuclear density. The incompressibility $K$, however, is less
well determined since in semiperipheral reactions the average density
probed up to 1 AGeV is too low.

\section*{ Acknowledgments }

One of us (A.B.L.) is grateful to P. Danielewicz, W. G. Lynch, M. B. Tsang
and G. D. Westfall for useful discussions during his stay at MSU, and 
to P. Danielewicz for pointing out the importance of a proper fragment 
selection in the comparison with fragment flow data.

\newpage

\begin{description}
\item[Table 1.] Parameter sets for the different mean fields employed in 
the BUU calculation.
\end{description}

\vspace{0.5cm}

\begin{center}
\begin{tabular}{|c|c|c|c|c|c|c|c|}
\hline
Not. & K (MeV) & $m^*/m_0$ & A (MeV) & B (MeV) & C (MeV) & 
$\tau$ & $\Lambda$ (fm$^{-1}$) \\
\hline
H  & 380 & 1.00 & -124.3 & 71.0  &  0.0  & 2.00 & -    \\
HM & 379 & 0.68 & -10.0  & 38.0  & -63.6 & 2.40 & 2.13 \\
HM1& 379 & 0.62 &  22.2  & 29.5  & -82.7 & 2.62 & 2.13 \\ 
SM & 220 & 0.68 & -108.6 & 136.8 & -63.6 & 1.26 & 2.13 \\
\hline
\end{tabular}
\end{center}

\newpage

\section*{ Figure captions }

\begin{description}

\item[Fig. 1] Density dependence of the energy per nucleon 
in nuclear matter (upper left panel). The other three panels show
the mean-field potential vs. momentum at the nuclear densities 
$0.5\rho_0$, $\rho_0$ and $2\rho_0$. The dashed, dotted, solid and
dash-dotted lines correspond to the interactions H, HM, HM1 and SM,
respectively.   

\item[Fig. 2] Central baryon density (smooth lines, left scale) and NN 
collision rate (histograms, right scale) vs. time for Au+Au at 0.4 AGeV 
and $b=6$ fm. The solid (dashed) line and histogram correspond the HM (SM) 
calculation. The horizontal dotted line shows the normal nuclear density 
$\rho_0$ for comparison. 

\item[Fig. 3] Average transverse velocity (a) and transverse 
momentum (b) of neutrons in the c.m. rapidity range $|y|<0.1$ 
emitted in the reaction plane ($\phi = -30^o \div 30^o,~ 150^o \div 180^o$,
$-180^o \div -150^o$) and out of the reaction plane
($\phi = 60^o \div 120^o,~ -60^o \div -120^o$) for Au+Au at 400 AMeV
and $b=6$ fm.

\item[Fig. 4] Time evolution of the baryon density in the reaction plane
$(x,z)$ for Au+Au at 0.4 AGeV and $b=6$ fm calculated with the HM mean
field. Isolines from outer to inner correspond to densities in the order 
0.05, 0.1, 0.15, 0.20 and 0.25 fm$^{-3}$. 

\item[Fig. 5] (a) Azimuthal dependence of the average neutron velocity
and (b) neutron azimuthal distribution for Au+Au at 0.4 AGeV and 
$b=5\div7$ fm. Neutrons are selected in the c.m. rapidity interval 
$|y|<0.1$. The corresponding elliptic flows are: $v_2=-0.046$ (H)
and $v_2=-0.090$ (HM).

\item[Fig. 6] The elliptic flow $v_2$ vs. transverse velocity 
(left panels) and vs. transverse momentum (right panels) of neutrons 
in the c.m. rapidity interval $|y|<0.1$ for the time steps
$t=30$ fm/c (upper panels) and $t=50$ fm/c (lower panels). The calculations
have been performed with the H (open circles, dashed line) and HM (full 
circles, solid line) parameter sets. The system is Au+Au at 0.4 AGeV and 
$b=6$ fm. 

\item[Fig. 7] Excitation function of the elliptic flow $v_2$ as 
given by the collection of experimental data from various groups 
\cite{Herrm99} and by the BUU calculations.
The BUU curves are shown for Au+Au collisions at $b=5 \div 7$ fm 
for {\it free} (unbound) protons with a critical distance $d_c=3$ fm 
in the c.m. rapidity range $|y|<0.1$ (see text). The curves correspond
to various mean-field parameter sets: short-dashed -- H, 
thick solid -- SM, dotted -- HM. For reference the results of 
the cascade calculation (without mean field) are shown by the thin 
solid line.

\item[Fig. 8] Excitation function of the transverse in-plane flow
$F$ for Au+Au collisions in comparison to the EOS \cite{Part95}
and E895 \cite{Liu00} proton data. The BUU curves are marked as 
in Fig. 7. The BUU results are impact parameter weighted in 
the interval $b=5 \div 7$ fm and are obtained for {\it free} (unbound) 
protons with a critical distance $d_c=3$ fm in the normalized c.m. 
rapidity range $-0.2 < y^{(0)} < 0.3$.

\item[Fig. 9] Excitation function of the transverse in-plane flow
$F$ for Au+Au collisions calculated with the SM mean field for
{\it free} protons ($d_c=3$ fm, dashed line)
and {\it all} protons ($d_c=0$ fm, solid line) 
in comparison to the experimental data
of the MSU-$4\pi$ Collaboration for $Z=2$ fragments \cite{Mag00}
and with the data of the EOS and FOPI Collaborations for $Z=1\div2$ 
fragments \cite{Liu98}. The BUU results are weighted in the impact 
parameter interval $b=5 \div 7$ fm.

\item[Fig. 10] Azimuthal distributions of neutrons with respect to the flow
axis $(\phi^\prime)$ for collisions of Nb+Nb and La+La at 0.4 AGeV
in comparison to the data (histograms) from Ref. \cite{Htun99}. 
Solid and dashed lines correspond to BUU calculations with HM1 and HM 
mean-field parametrizations, respectively.
The BUU results are weighted
in the impact parameter range $b=1\div5$ fm (Nb+Nb) and $b=1\div6$ fm
(La+La) according to the estimate $b_{max}=0.5(R_p+R_t)$ in \cite{Htun99}.
In the calculations we selected {\it free} neutrons ($d_c=3$ fm)
with longitudinal momenta $-0.1 \leq (p_z^\prime/p_{proj}^\prime)_{c.m.} 
\leq 0.1$.

\item[Fig. 11] Average neutron transverse in-plane momentum vs. normalized 
c.m. rapidity for Nb+Nb and La+La collisions at 0.4 AGeV in comparison 
to the data from \cite{Htun99}. The BUU results are weighted in the impact 
parameter range $b=1\div5$ fm (Nb+Nb) and $b=1\div6$ fm (La+La); free neutrons
with a critical distance $d_c=3$ fm are selected in the BUU calculations.
The solid and dashed lines correspond to the HM1 and HM parametrizations.
The experimental threshold laboratory kin. energy of 55 MeV for neutrons 
\cite{Htun99} is taken into account which leads to the asymmetric 
$\langle p_x \rangle (y^{(0)})$-dependence with respect to $y^{(0)}=0$.

\item[Fig. 12] The neutron squeeze-out ratio $R_N$ vs. impact parameter $b$ 
for Au+Au collisions at 400 AMeV. The data from Ref. \cite{Lambr94} are 
for neutrons in the rapidity interval $0.4 \leq y/y_{proj} < 0.6$,
while the histograms represent the BUU calculations with various mean-field 
parameter sets. Upper panel: short-dashed -- H, dash-dotted -- SM. 
Lower panel: dotted -- HM, solid -- HM1. 
In the BUU calculations free neutrons are selected with $d_c=3$ fm and 
the angular cuts of the LAND detector are taken into account. The calculated 
azimuthal distributions are shown with respect to the reaction plane given by 
the ${\bf Q}$ vector (see text).

\item[Fig. 13] Squeeze-out ratio of neutrons $R_N$ in the rapidity 
interval $0.4 \leq y/y_{proj} < 0.6$ as a function of the neutron
transverse momentum for collisions of Au+Au at 400, 600 and 800 AMeV.
The data are from Ref. \cite{Lambr94}; the histograms show the BUU 
calculations for different mean fields: short-dashed -- H, dotted -- HM, 
solid -- HM1, dash-dotted -- SM, long-dashed -- HM1 with reduced (by 30\%) 
NN cross section. The BUU results are impact parameter weighted in the region
$b=5\div9$ fm, which approximately corresponds to the centrality bin E2 for 
the data. In the calculations the reaction plane was determined by the 
${\bf Q}$ vector.

\end{description}

\clearpage
\thispagestyle{empty}

\begin{figure}[btp]
\psfig{figure=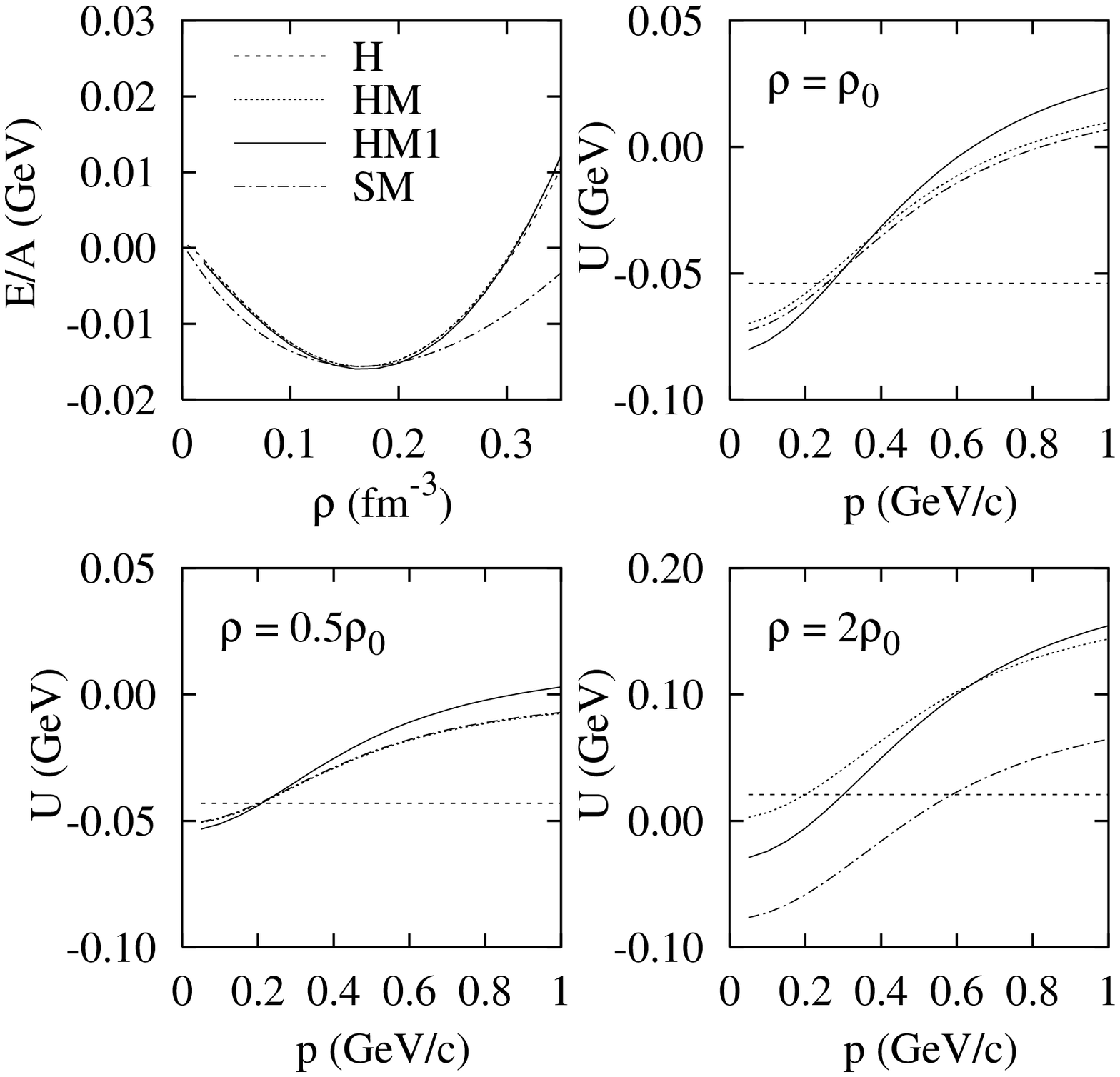,width=0.8\textwidth}
\caption{ }
\end{figure}

\clearpage
\thispagestyle{empty}

\begin{figure}[btp]
\psfig{figure=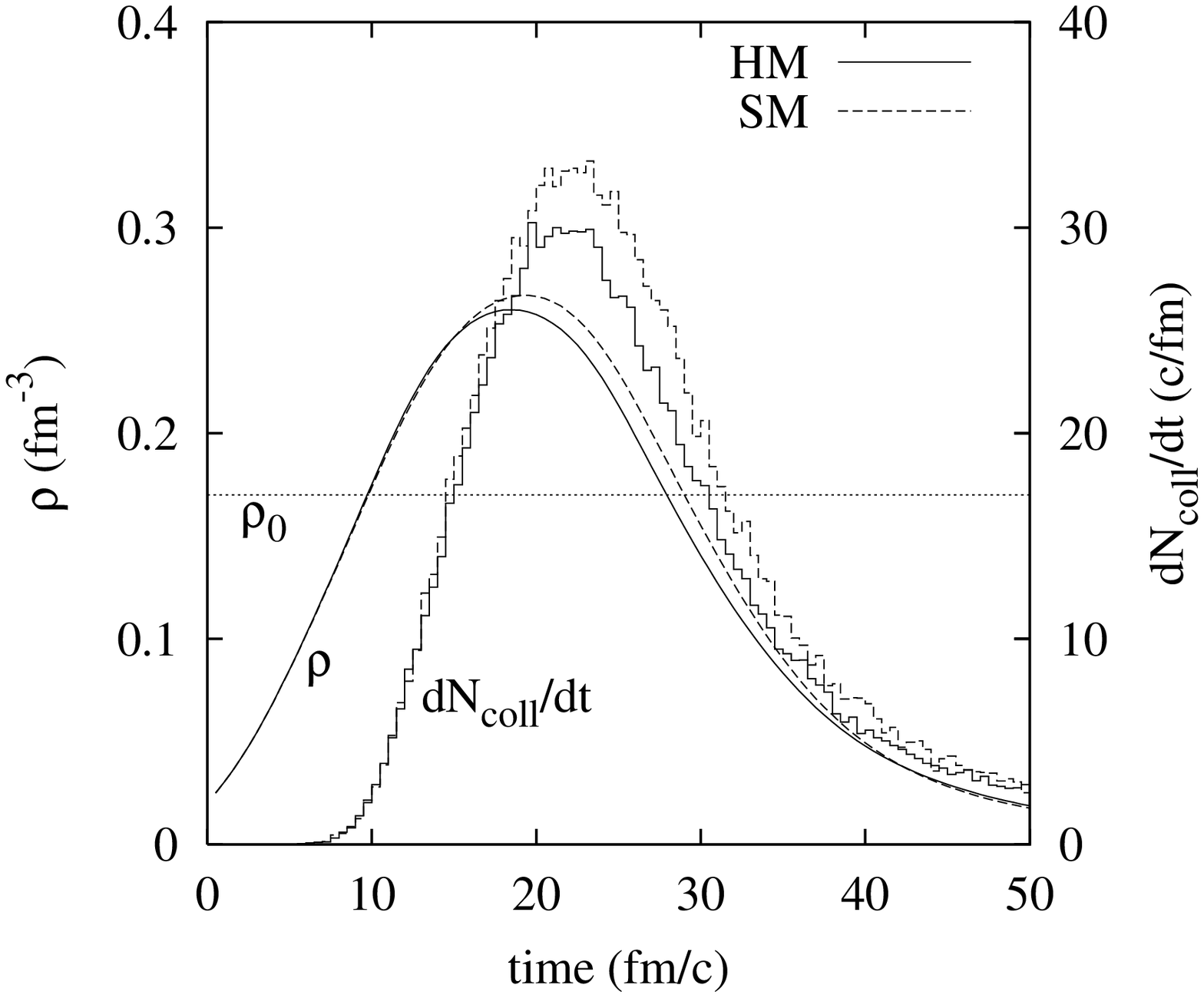,width=0.8\textwidth}
\caption{ }
\end{figure}

\clearpage
\thispagestyle{empty}

\begin{figure}[btp]
\psfig{figure=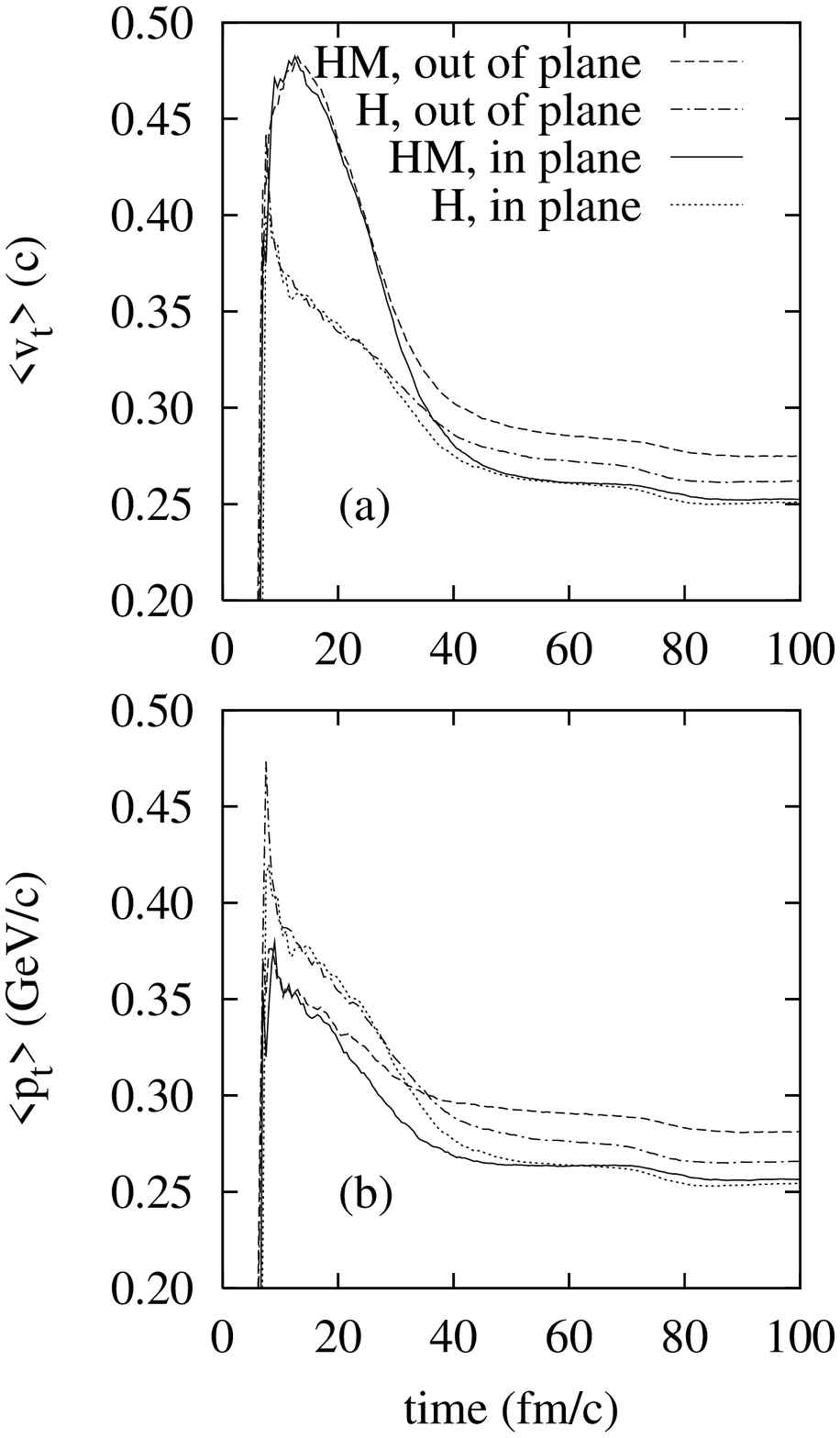,width=0.8\textwidth}
\caption{ }
\end{figure}

\clearpage
\thispagestyle{empty}

\begin{figure}[btp]
\psfig{figure=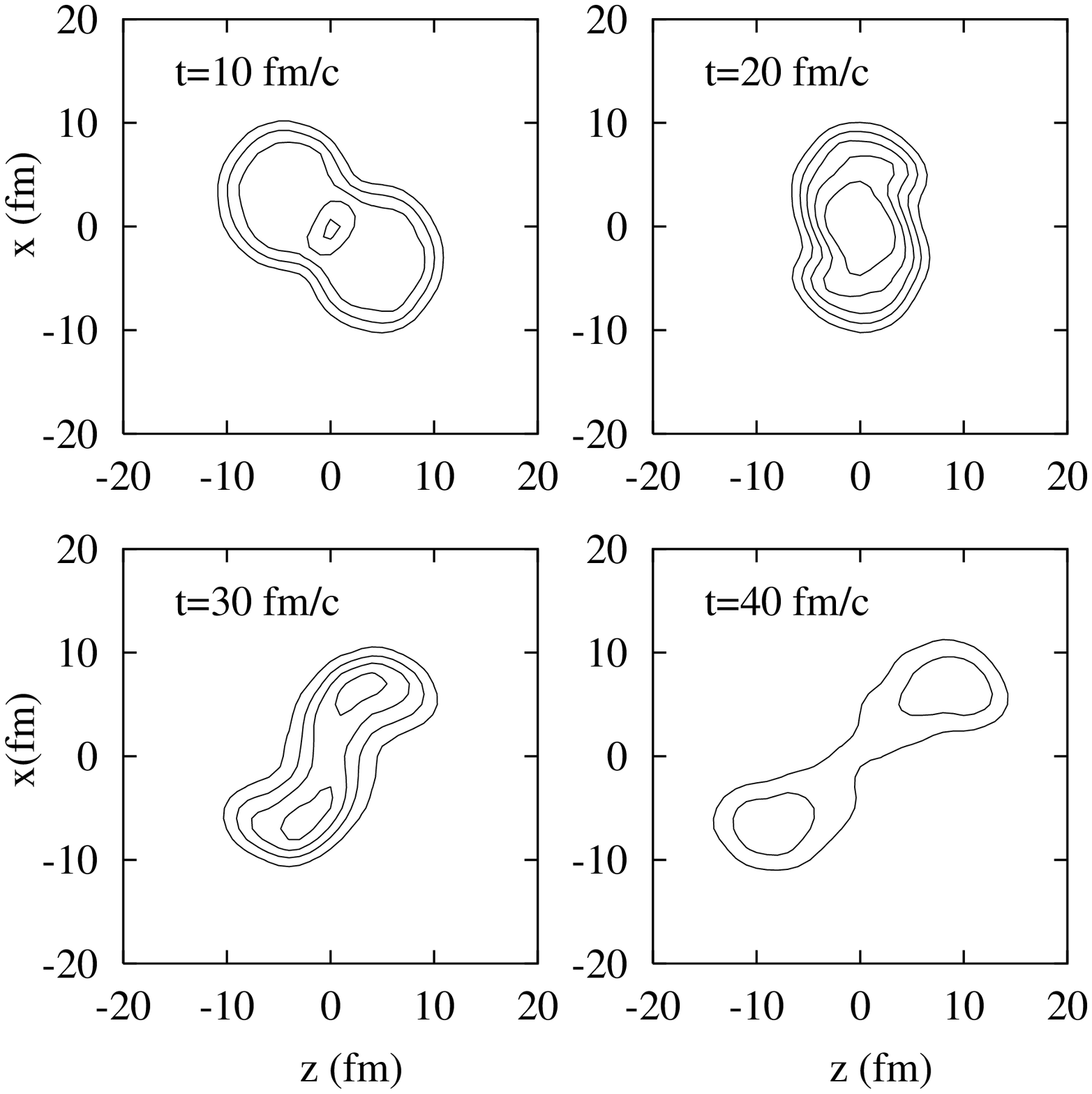,width=0.8\textwidth}
\caption{ }
\end{figure}

\clearpage
\thispagestyle{empty}

\begin{figure}[btp]
\psfig{figure=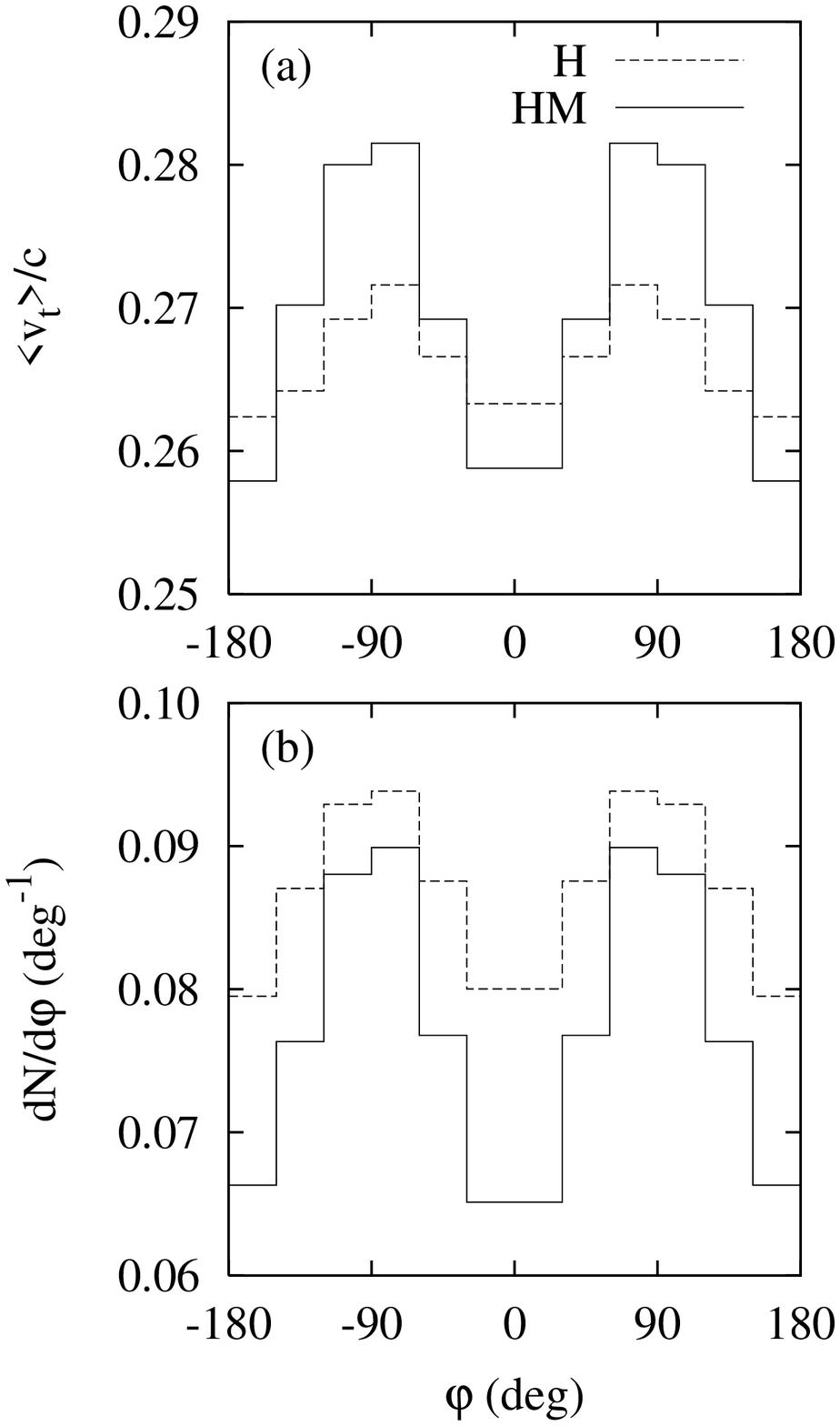,width=0.8\textwidth}
\caption{ }
\end{figure}

\clearpage
\thispagestyle{empty}

\begin{figure}[btp]
\psfig{figure=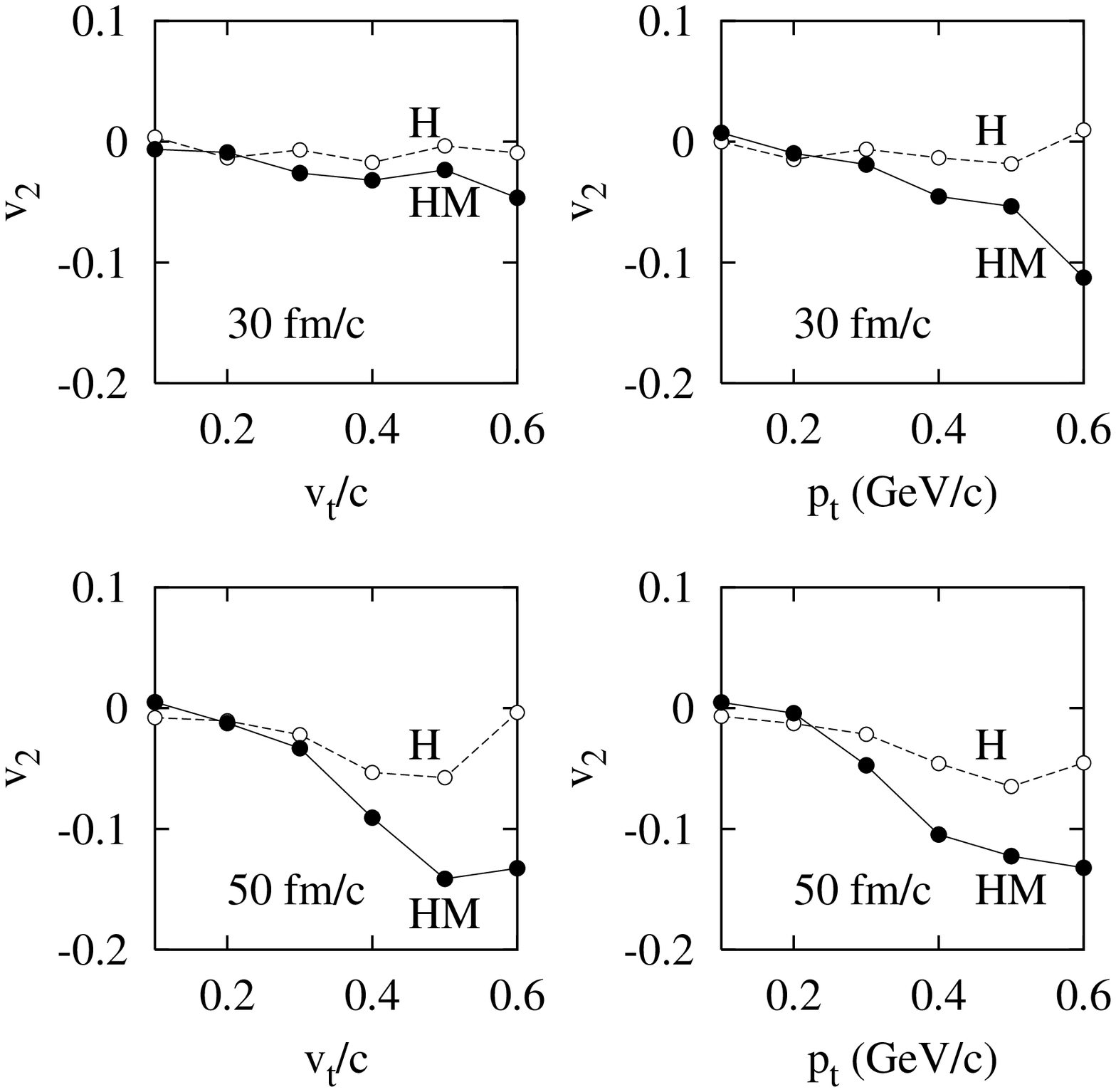,width=0.8\textwidth}
\caption{ }
\end{figure}

\clearpage
\thispagestyle{empty}

\begin{figure}[btp]
\psfig{figure=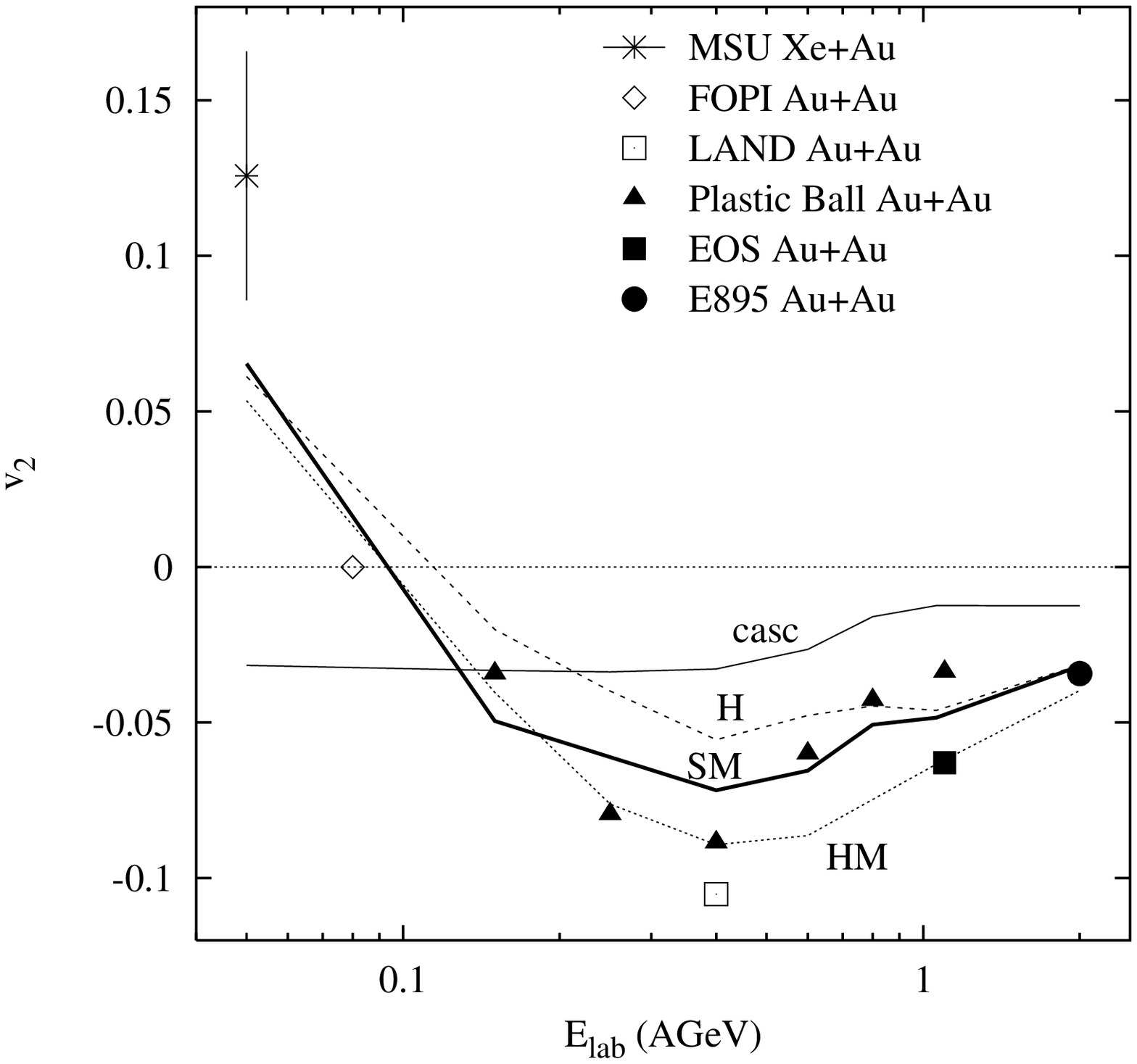,width=0.8\textwidth}
\caption{ }
\end{figure}

\clearpage
\thispagestyle{empty}

\begin{figure}[btp]
\psfig{figure=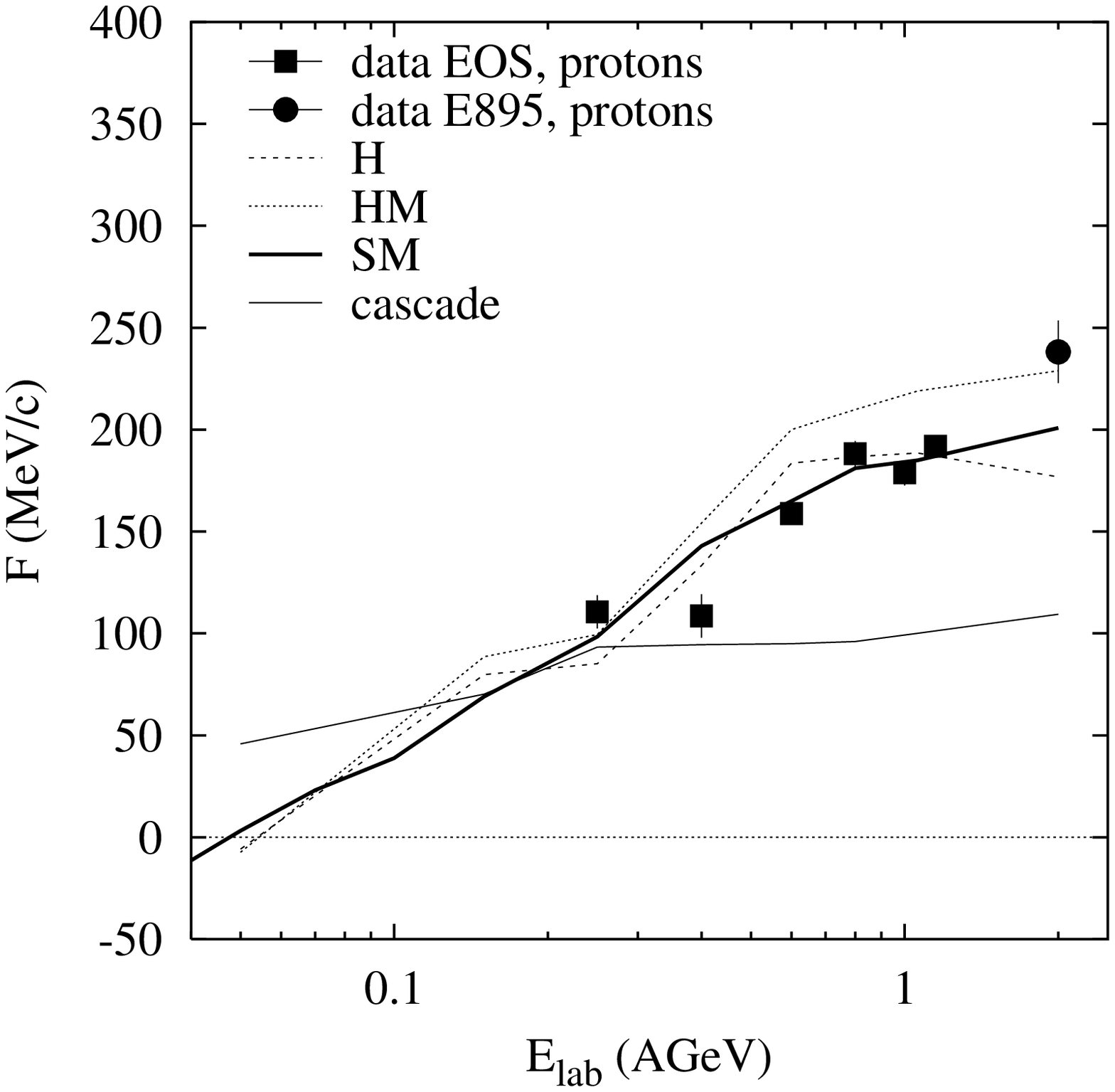,width=0.8\textwidth}
\caption{ }
\end{figure}

\clearpage
\thispagestyle{empty}

\begin{figure}[btp]
\psfig{figure=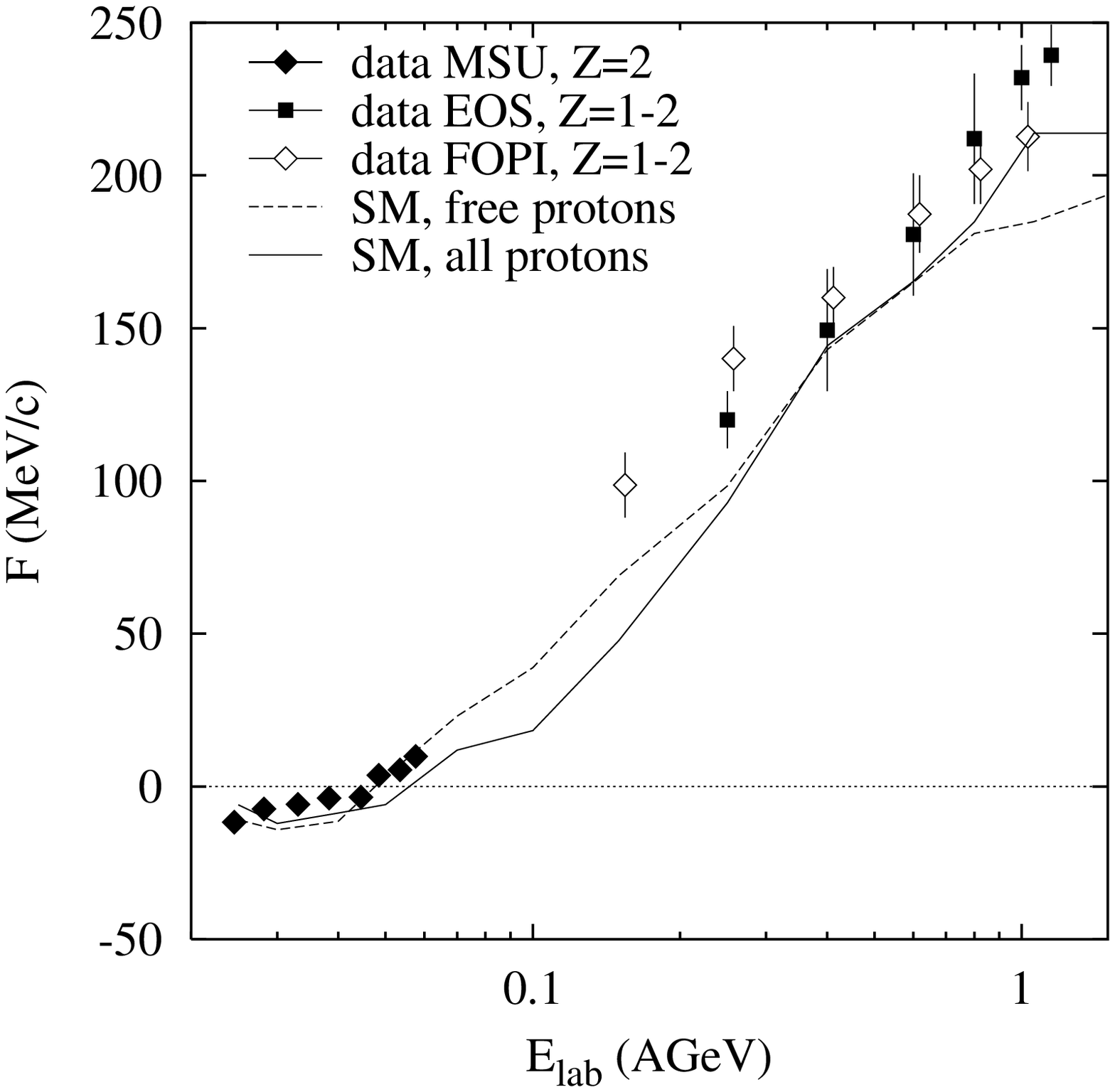,width=0.8\textwidth}
\caption{ }
\end{figure}

\clearpage
\thispagestyle{empty}

\begin{figure}[btp]
\psfig{figure=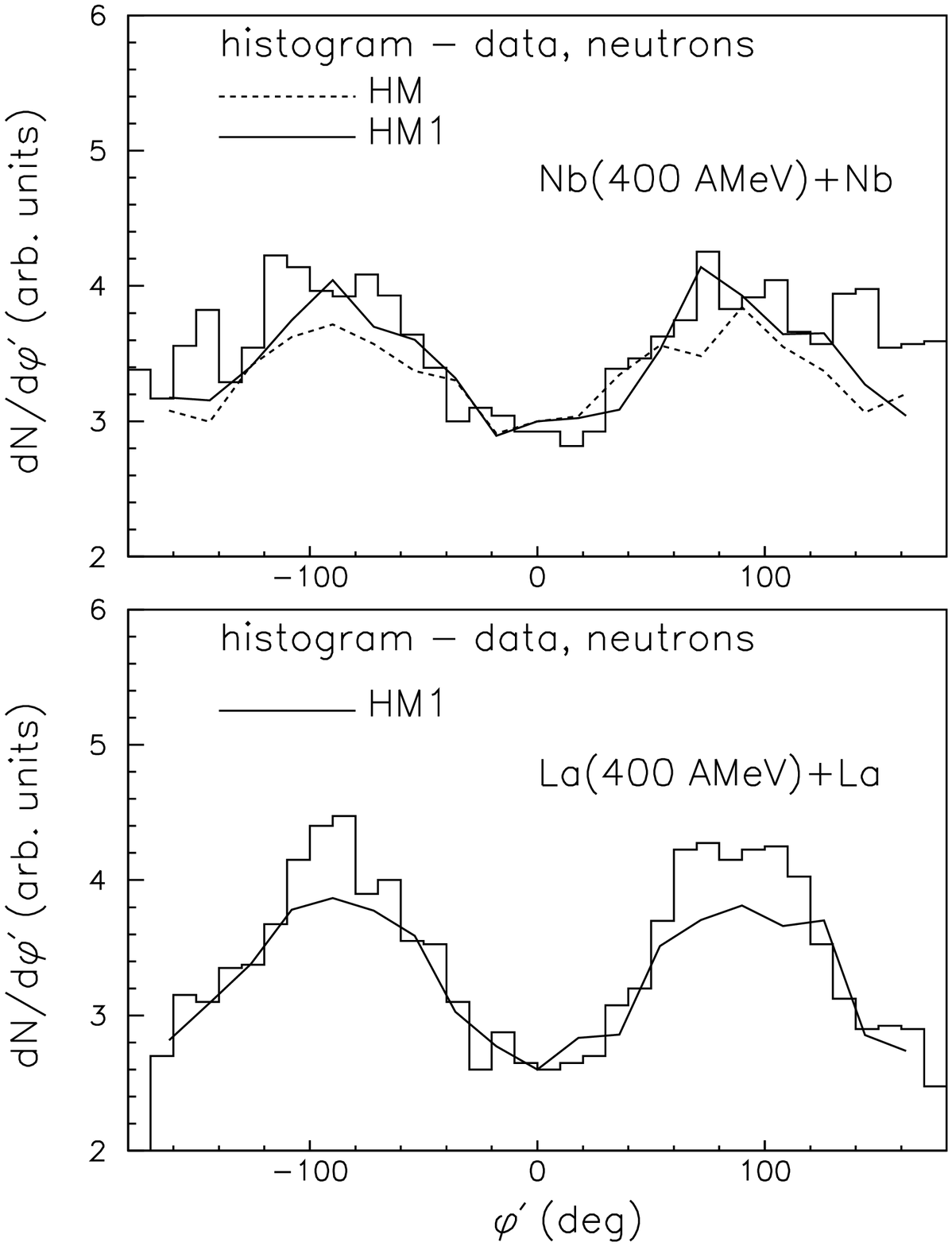,width=0.8\textwidth}
\caption{ }
\end{figure}

\clearpage
\thispagestyle{empty}

\begin{figure}[btp]
\psfig{figure=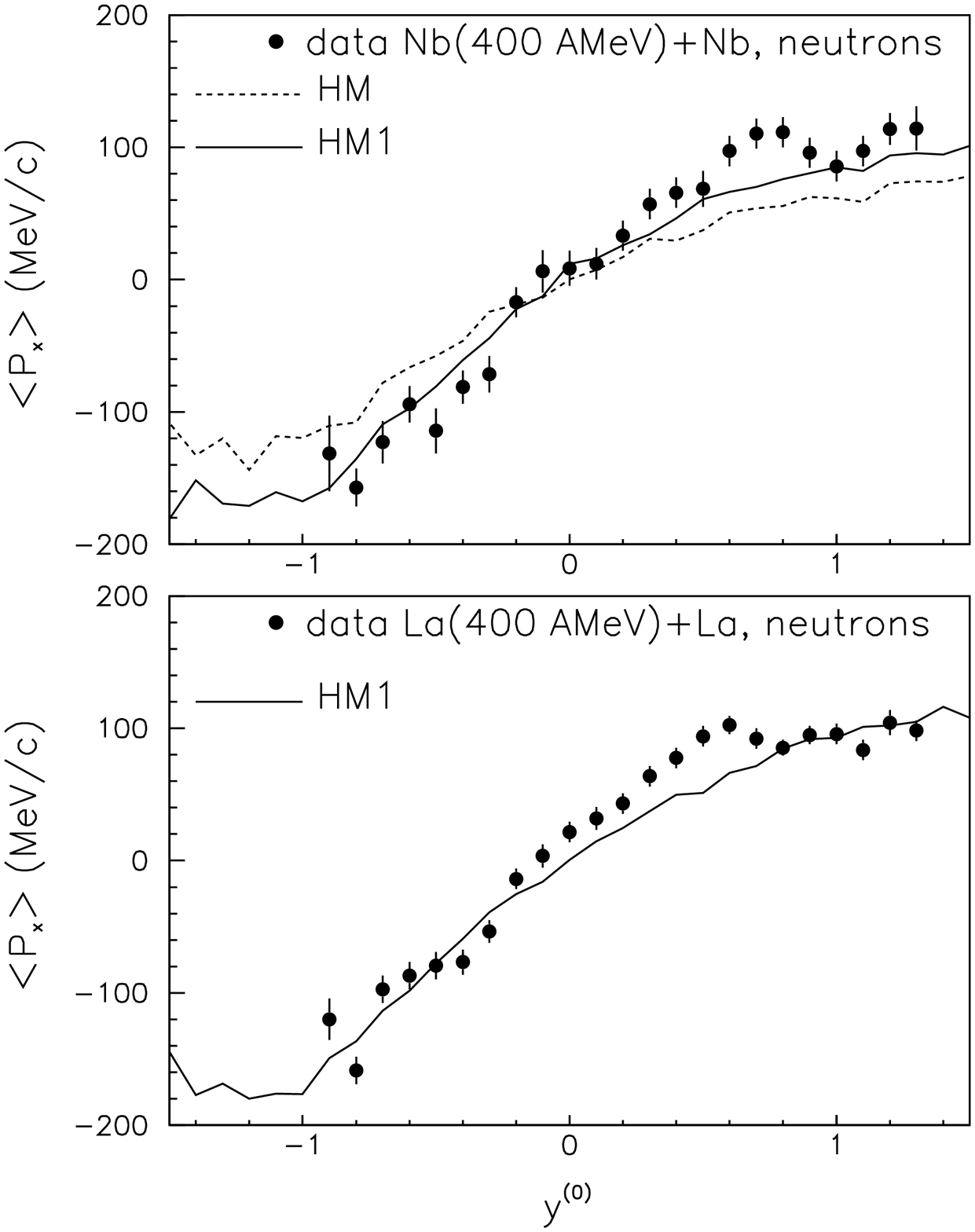,width=0.8\textwidth}
\caption{ }
\end{figure}

\clearpage
\thispagestyle{empty}

\begin{figure}[btp]
\psfig{figure=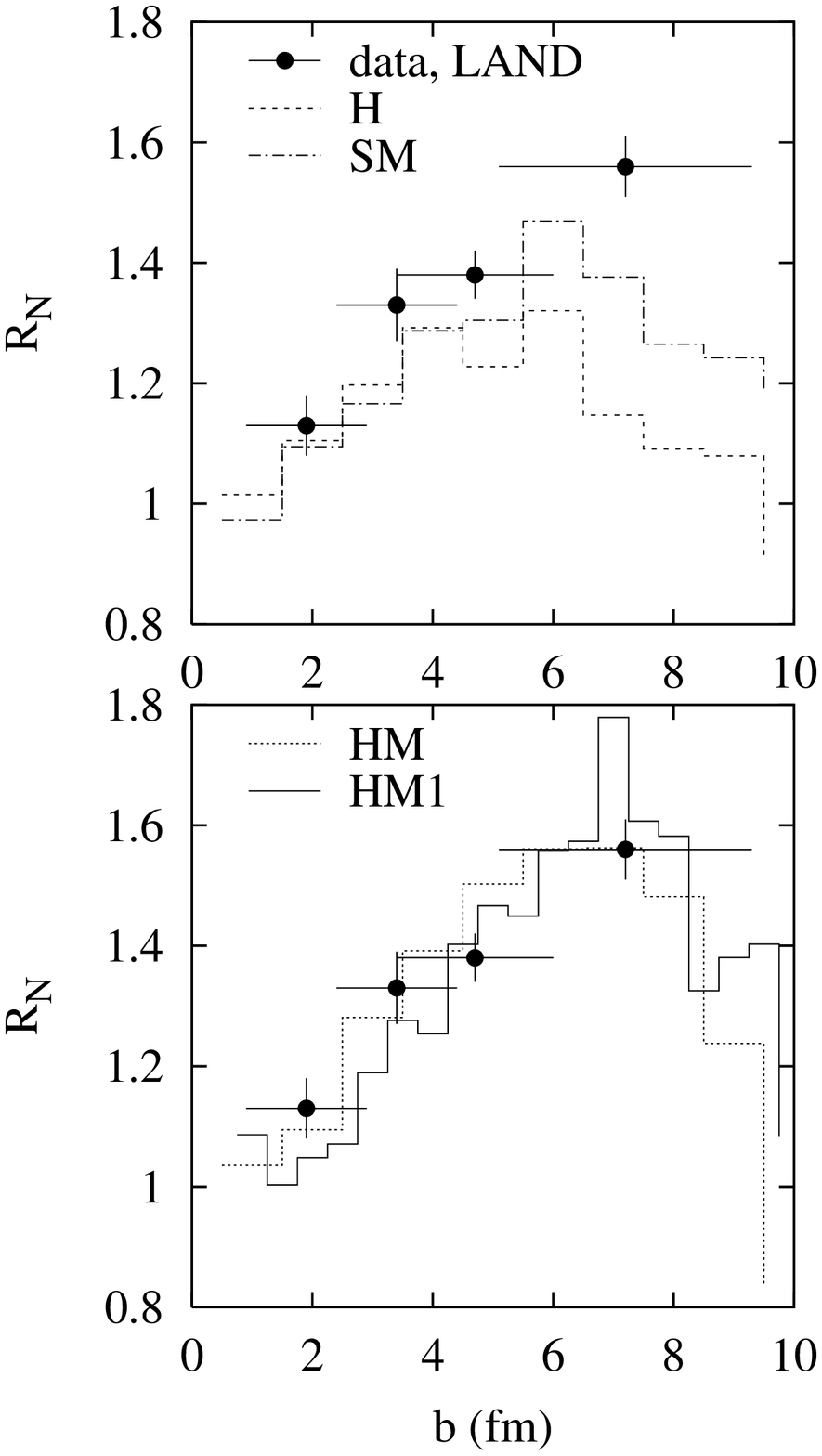,width=0.8\textwidth}
\caption{ }
\end{figure}

\clearpage
\thispagestyle{empty}

\begin{figure}[btp]
\psfig{figure=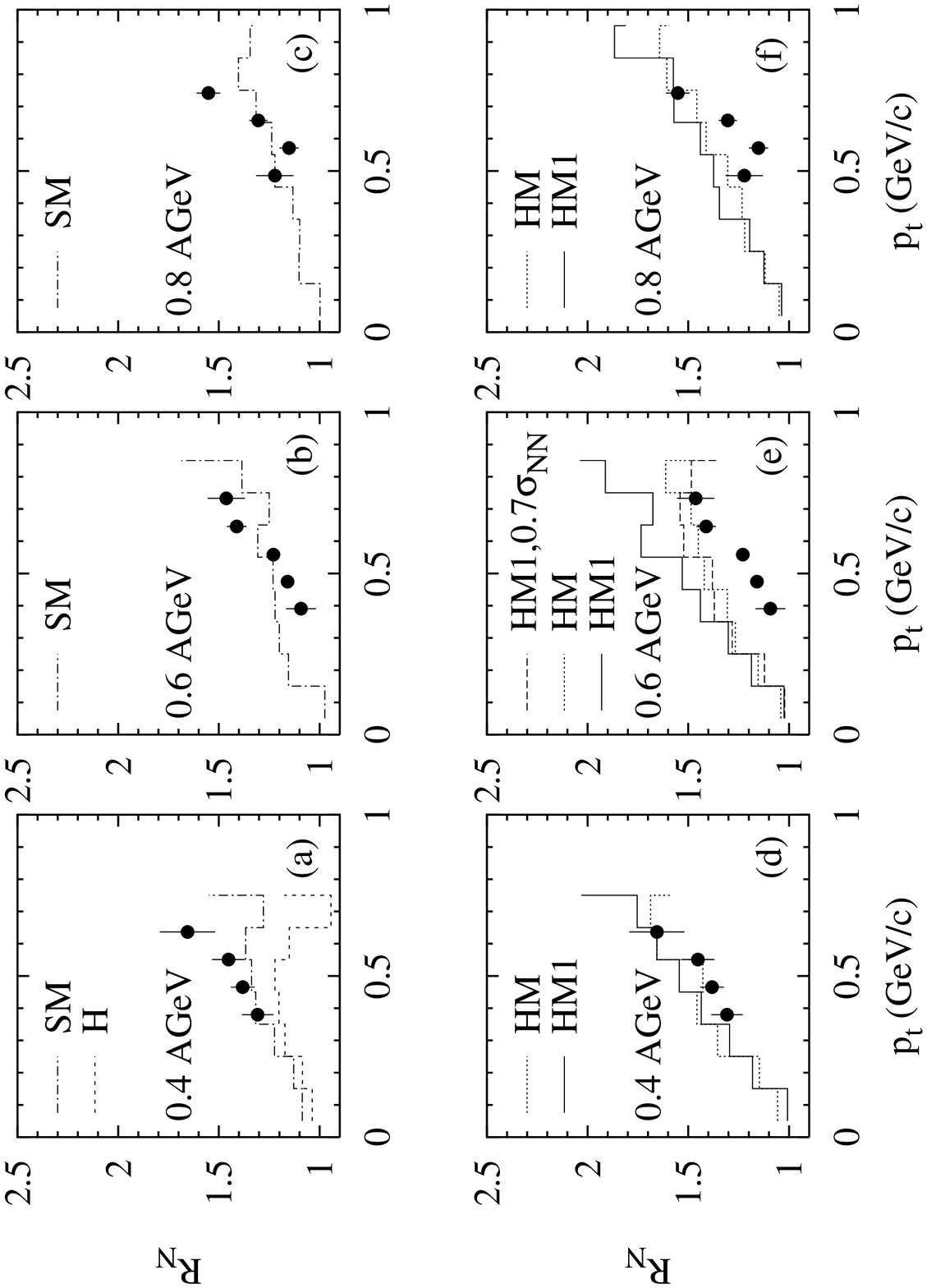,width=0.8\textwidth}
\caption{ }
\end{figure}


\newpage
    
\begin{thebibliography}{99}

\bibitem{Scheid68} W. Scheid, R. Ligensa, and W. Greiner,
Phys. Rev. Lett. {\bf 21}, 1479 (1968).

\bibitem{SG86} H. St\"ocker and W. Greiner, 
Phys. Rep. {\bf 137}, 277 (1986).

\bibitem{BD88} G.F. Bertsch and S. Das Gupta, 
Phys. Rep. {\bf 160}, 189 (1988).

\bibitem{Gut89} H.H. Gutbrod, A.M. Poskanzer, and H.G. Ritter,
Rep. Prog. Phys. {\bf 52}, 1267 (1989).

\bibitem{Aich91} J. Aichelin, Phys. Rep. {\bf 202}, 233 (1991).

\bibitem{Reis97} W. Reisdorf and H.G. Ritter, 
Annu. Rev. Nucl. Part. Sci. {\bf 47}, 1 (1997).

\bibitem{Sahu98} P.K. Sahu, A. Hombach, W. Cassing, M. Effenberger,
U. Mosel, Nucl. Phys. {\bf A640}, 493 (1998).

\bibitem{Herrm99} N. Herrmann, J.P. Wessels, and T. Wienold,
Annu. Rev. Nucl. Part. Sci. {\bf 49}, 581 (1999).

\bibitem{Hombach99} A. Hombach, W. Cassing, S. Teis and U. Mosel,
Eur. Phys. J. A {\bf 5}, 157 (1999).

\bibitem{GBD87} C. Gale, G. Bertsch, S. Das Gupta, 
Phys. Rev. C {\bf 35}, 1666 (1987).

\bibitem{Aich87} J. Aichelin, A. Rosenhauer, G. Peilert,
H. St\"ocker, and W. Greiner, Phys. Rev. Lett. {\bf 58}, 1926 (1987).

\bibitem{Gale90} C. Gale, G.M. Welke, M. Prakash, S.J. Lee,
and S. Das Gupta, Phys. Rev. C {\bf 41}, 1545 (1990).

\bibitem{Zhang94} J. Zhang, S. Das Gupta, and C. Gale,
Phys. Rev. C {\bf 50}, 1617 (1994).

\bibitem{DO85} P. Danielewicz and G. Odyniec, 
Phys. Lett. {\bf 157B}, 146 (1985).

\bibitem{balance} H.M. Xu, Phys. Rev. Lett. {\bf 67}, 2769 (1992);
Phys. Rev. C {\bf 46}, R392 (1992).

\bibitem{Victor} V.N. Russkikh, Yu.B. Ivanov, Yu.E. Pokrovsky,
P.A. Henning, Nucl. Phys. {\bf A572}, 749 (1994). 

\bibitem{Homb99} A. Hombach, W. Cassing, and U. Mosel,
Eur. Phys. J. A {\bf 5}, 77 (1999).

\bibitem{Dan99} P. Danielewicz, Nucl. Phys. {\bf A673}, 375 (2000).

\bibitem{Brill96} D. Brill, P. Beckerle, C. Bormann, E. Schwab,
Y. Shin, R. Stock, H. Str\"obele, P. Baltes, C. M\"untz, H. Oeschler,
C. Sturm, A. Wagner, R. Barth, M. Cie\'slak, M. D\c{e}bowski,
E. Grosse, P. Koczo\'n, M. Mang, D. Mi\'skowiec, R. Schicker,
P. Senger, B. Kohlmeyer, F. P\"uhlhofer, J. Speer, K. V\"olkel,
W. Walu\'s, Z. Phys. A {\bf 355}, 61 (1996).

\bibitem{Sahu2000} P.K. Sahu, W. Cassing, U. Mosel and A. Ohnishi,
Nucl. Phys. {\bf A672}, 376 (2000).

\bibitem{EBM99} M. Effenberger, E.L. Bratkovskaya, and U. Mosel,
Phys. Rev. C {\bf 60}, 044614 (1999).

\bibitem{Cugnon96} J. Cugnon, D. L'H\^ote, J. Vandermeulen,
Nucl. Instrum. Methods Phys. Res. B {\bf 111}, 215 (1996).

\bibitem{Teis97} S. Teis, W. Cassing, M. Effenberger, A. Hombach,
U. Mosel, and G. Wolf, Z. Phys. A {\bf 356}, 421 (1997).

\bibitem{Welke88} G.M. Welke, M. Prakash, T.T.S. Kuo, S. Das Gupta,
and C. Gale, Phys. Rev. C {\bf 38}, 2101 (1988).

\bibitem{Prakash97} Madappa Prakash, Ignazio Bombaci, Manju Prakash,
Paul J. Ellis, James M. Lattimer, Roland Knorren, 
Phys. Rep. {\bf 280}, 1 (1997).

\bibitem{Chab97} E. Chabanat, P. Bonche, P. Haensel, J. Meyer,
R. Schaeffer, Nucl. Phys. {\bf A627}, 710 (1997).

\bibitem{App98} H. Appelsh\"auser et al. (NA49 Collaboration),
Phys. Rev. Lett. {\bf 80}, 4136 (1998).

\bibitem{E895} C. Pinkenburg et al. (E895 Collaboration),
Phys. Rev. Lett. {\bf 83}, 1295 (1999).

\bibitem{Gosset77} J. Gosset, H.H. Gutbrod, W.G. Meyer, 
A.M. Poskanzer, A. Sandoval, R. Stock, and G.D. Westfall, 
Phys. Rev. C {\bf 16}, 629 (1977).

\bibitem{Tsang96} M.B. Tsang, P. Danielewicz, W.C. Hsi, M. Huang,
W.G. Lynch, D.R. Bowman, C.K. Gelbke, M.A. Lisa, G.F. Peaslee,
R.J. Charity, L.G. Sobotka, and ALADIN Collaboration,
Phys. Rev. C {\bf 53}, 1959 (1996).

\bibitem{Rai99} G. Rai et al., Nucl. Phys. {\bf A661}, 162c (1999). 

\bibitem{Doss86} K.G.R. Doss, H.\AA. Gustafsson, H.H. Gutbrod,
K.H. Kampert, B. Kolb, H. L\"ohner, B. Ludewigt, A.M. Poskanzer,
H.G. Ritter, H.R. Schmidt, and H. Wieman,
Phys. Rev. Lett. {\bf 57}, 302 (1986).

\bibitem{Zheng99} Yu-Ming Zheng, C.M. Ko, Bao-An Li, and Bin Zhang,
Phys. Rev. Lett. {\bf 83}, 2534 (1999).

\bibitem{Soff} S. Soff, S.A. Bass, C. Hartnack, H. St\"ocker 
and W. Greiner, Phys. Rev. C {\bf 51}, 3320 (1995).

\bibitem{Part95} M.D. Partlan et al. (EOS Collaboration),
Phys. Rev. Lett. {\bf 75}, 2100 (1995).

\bibitem{Liu98} H. Liu et al. (E895 Collaboration), 
Nucl. Phys. {\bf A638}, 451c (1998).
 
\bibitem{Liu00} H. Liu et al. (E895 Collaboration), 
Phys. Rev. Lett. {\bf 84}, 5488 (2000).

\bibitem{Mag00} D.J. Magestro, W. Bauer, O. Bjarki, J.D. Crispin,
M.L. Miller, M.B. Tonjes, A.M. Vander Molen, G.D. Westfall,
R. Pak, E. Norbeck, Phys. Rev. C {\bf 61}, 021602(R) (2000).

\bibitem{Crochet97} P. Crochet et al. (FOPI Collaboration),
Nucl. Phys. {\bf A624}, 755 (1997).

\bibitem{Htun99} M.M. Htun, R. Madey, W.M. Zhang, M. Elaasar,
D. Keane, B.D. Anderson, A.R. Baldwin, J. Jiang, A. Scott,
Y. Shao, J.W. Watson, K. Frankel, L. Heilbronn, G. Krebs,
M.A. McMahan, W. Rathbun, J. Schambach, G.D. Westfall, S. Yennello,
C. Gale, and J. Zhang, Phys. Rev. C {\bf 59}, 336 (1999).

\bibitem{Gy82} M. Gyulassy, K.A. Frankel and H. St\"ocker,
Phys. Lett. {\bf 110B}, 185 (1982).

\bibitem{Gutbrod89} H.H. Gutbrod, K.H. Kampert, B.W. Kolb, 
A.M. Poskanzer, H.G. Ritter and H.R. Schmidt, 
Phys. Lett. {\bf 216B}, 267 (1989).

\bibitem{Gutbrod90} H.H. Gutbrod, K.H. Kampert, B. Kolb, 
A.M. Poskanzer, H.G. Ritter, R. Schicker, and H.R. Schmidt, 
Phys. Rev. C {\bf 42}, 640 (1990).

\bibitem{Lambr94} D. Lambrecht, T. Blaich, T.W. Elze, H. Emling,
H. Freiesleben, K. Grimm, W. Henning, R. Holzmann, J.G. Keller,
H. Klingler, J.V. Kratz, R. Kulessa, S. Lange, Y. Leifels,
E. Lubkiewicz, E.F. Moore, W. Prokopowicz, R. Schmidt, C. Sch\"utter,
H. Spies, K. Stelzer, J. Stroth, E. Wajda, W. Walu\'s, M. Zinser,
E. Zude, and FOPI Collaboration,  
Z. Phys. A {\bf 350}, 115 (1994).

\end{thebibliography}
\end{document}